# A Novel Quantum-Classical Hybrid Algorithm for Determining Eigenstate Energies in Quantum Chemistry

Qing-Xing Xie[a*], Yan Zhao[bc**]

[a]*Department of Physics, Hubei University, Wuhan 430062, P. R. China.*

[b] *College of Materials Science and Engineering, Sichuan University, Chengdu 610065, P. R. China.*

[c]*The Institute of Technological Sciences, Wuhan University, Wuhan 430072, P. R. China.*

[*]xieqx@hubu.edu.cn [**]yanzhao@scu.edu.cn

## ABSTRACT

Developing efficient quantum computing algorithms is crucial for addressing computationally challenging problems across various fields. In this paper, we present a novel quantum algorithm, denoted as XZ24, designed for the efficient computation of the eigen-energy spectra of arbitrary quantum systems. Given a Hamiltonian $\hat{H}$ and an arbitrary initial reference state $|\psi_{\text{ref}}\rangle$, the algorithm extracts information about $\langle\psi_{\text{ref}}|\cos(\hat{H}t)|\psi_{\text{ref}}\rangle$ from the state of an auxiliary qubit. By applying a Fourier transform to this signal, the algorithm resolves the energies of all eigenstates of the Hamiltonian that have non-negligible overlap with the reference wavefunction. We provide a comprehensive theoretical analysis alongside numerical simulations, demonstrating the algorithm's superior computational efficiency and accuracy. Compared to existing quantum algorithms, XZ24 offers three significant advantages. First, it eliminates the need for eigenstate preparation on the quantum processor, requiring only a reference state with non-negligible overlap with the target eigenstate. This is a significant improvement over methods such as the Variational Quantum Eigensolver and single-qubit variants of Quantum Phase Estimation, which demand substantial resources for eigenstate preparation. Second, XZ24 significantly reduces measurement overhead, as only a single auxiliary qubit is measured during execution. For a quantum system of size $L$, with an absolute precision $\varepsilon$, the sampling complexity scales linearly with the system size as $O(L \cdot \varepsilon^{-1})$. When only relative precision $\epsilon$ is required, the sampling complexity scales as $O(\epsilon^{-1})$, making the number of measurements independent of system size. Third, this algorithm facilitates the efficient simultaneous computation of multiple eigen-energies, depending on the choice of reference state. We expect that this algorithm will catalyze substantial advancements in quantum system simulations and offer promising applications in quantum computing and quantum simulations.

## 1. Introduction

Quantum computing has emerged as a transformative computational paradigm, utilizing principles of quantum mechanics to tackle challenges that exceed the capabilities of classical computers across diverse domains [1–9]. A notable challenge lies in precisely simulating intricate quantum systems, a task that poses difficulties for classical computing architectures. Nonetheless, by leveraging the unique characteristic of quantum superposition and entanglement, qubits exhibit theoretical potential in representing complex quantum states, positioning quantum computing as an exceptionally promising avenue for this purpose [10–13]. In this thriving field, a key focus involves developing efficient algorithms that can accurately calculate eigenstates of quantum systems. These algorithms are essential for addressing computationally challenging problems spanning fields such as materials science, chemistry, and condensed matter physics [14–21].

Various quantum algorithms have been proposed, including Quantum Phase Estimation (QPE) [22–26], variational quantum eigensolver (VQE) [27–30], adiabatic state preparation (ASP) [31–35], and nonunitary time evolution (NTE) [36–39]. The QPE method requires a significant number of Hamiltonian evolution operators, introducing heightened challenges for circuit depth and fault tolerance in quantum hardware [22,24]. Similarly, the ASP method involves slow and prolonged system evolution, significantly escalating the quantum circuit cost [34,40]. The computational requirements of both methods exceed the capabilities of near-term quantum computers. While VQE has shown marked enhancements in quantum resource utilization compared to QPE and ASP, it encounters obstacles in ansatz circuit design [41]. The development of ansatz circuits that can efficiently prepare target quantum states remains an unresolved challenge [42–45]. Moreover, the VQE method entails the problems related to intricate nonlinear and non-convex optimizations of circuit parameters [46–49]. The recently introduced NTE method theoretically addresses prior limitations. Nevertheless, given that all quantum gates in digital quantum computers are unitary, except for the measurement gate, implementing this method encounters the hurdle of realizing non-unitary operator operations in quantum circuits [50–54].

In this paper, we introduce a novel non-variational quantum algorithm, referred to as XZ24 (Xie-Zhao 2024), designed for the efficient computation of eigen-energy spectra in arbitrary quantum systems. Unlike the VQE method, the XZ24 algorithm circumvents the challenges associated with ansatz circuit construction and complex ansatz circuit parameter optimization. The primary cost in quantum circuit for XZ24 arises from its reliance on a single-qubit controlled real-time Hamiltonian evolution operator ($e^{-i\hat{H}t}$), which significantly reduces the circuit depth compared to QPE that requires multiple controlled $e^{-i\hat{H}t}$ operators [22,24]. At first glance, the circuit complexity of the XZ24 algorithm appears comparable to that of the single-qubit variant of QPE, specifically the iterative QPE (IQPE) [26,55]. However, a key distinction lies in the nature of the input state. IQPE is designed to compute the eigenvalue of a given input eigenstate via phase estimation, which necessitates that the input state be an eigenstate of the Hamiltonian. This requirement introduces a fundamental difficulty in computational cost compared to standard QPE, where the input state need only have a non-zero overlap with the target eigenstate—typically a more easily achievable condition. In contrast, preparing an exact eigenstate in the quantum register, as required by IQPE, is generally more demanding. The XZ24 algorithm, akin to QPE, only requires a non-negligible overlap between the input state and the desired eigenstate, making it more practical than IQPE for real-world applications. Additionally, compared to the ASP algorithm, XZ24 does not require the long evolution times, thereby substantially reducing the circuit depth. As compared to the NTE algorithms, which struggle with the implementation of non-unitary operators using unitary quantum gates [39], all operators utilized in the XZ24 algorithm are unitary, thus avoiding these complexities.

In addition to the advantages mentioned above, XZ24 provides two further benefits. One significant benefit is the extremely low measurement cost. Compared to the VQE method, where the Hamiltonian is decomposed into numerous Pauli strings and the expectation value of each Pauli string is measured to derive the energy of the output quantum state, the XZ24 algorithm—similar to the IQPE method—requires sampling only a single ancillary qubit in the circuit during each step. Consequently, the sampling cost in per evolution-time sampling step does not increase with the size of the computational system. Furthermore, the XZ24 method can obtain multiple eigenstates simultaneously in a single execution; the specific eigenstates acquired depend on the chosen reference state. Thus, ground state and multiple lower-order excited states can be calculated in a single practical execution. In contrast, many methods for calculating excited states, such as OSRVE [56] and VQD [57], can only compute one eigenstate at a time, necessitating multiple computations to obtain these eigenstates [58,59]. As a result, the computational efficiency of XZ24 significantly surpasses that of existing methods. The rest of this paper is structured as follows. Section 2 details the XZ24 algorithm and discusses its implementation on quantum computers. Section 3 presents numerical simulations of molecular systems and condensed matter models to demonstrate the effectiveness of the XZ24 algorithm. We also provide a thorough analysis of the computational cost, particularly sampling complexity, highlighting the superiority of the XZ24 method over existing approaches. Finally, Section 4 concludes the paper.

## 2. Theory and Methodology

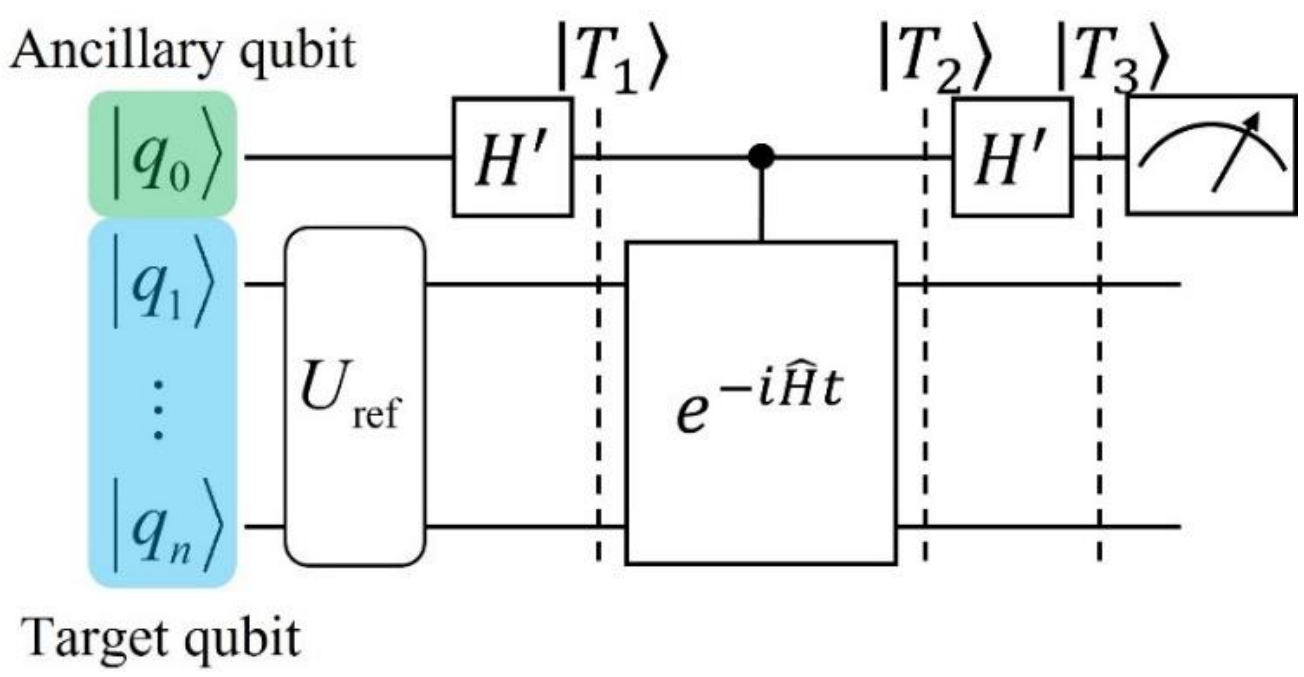

Figure 1. Quantum circuit for the XZ24 algorithm. The $q_0$ (highlighted within the green box) serves as the ancillary qubit, while qubits $q_1 \cdots q_n$ (enclosed in the cyan box) are designated as target qubits.

The quantum circuit structure of XZ24 is illustrated in Figure 1. An ancillary qubit, $q_0$, serves as the control qubit. The quantum circuit $U_{\text{ref}}$ is executed on target qubits $q_1 \cdots q_n$ to prepare the appropriate reference state wavefunction $|\psi_{\text{ref}}\rangle$. The choice of the reference state in the XZ24 algorithm is flexible, as long as the reference state has a non-zero wavefunction overlap with the target eigenstate. For ground-state energy calculations in atomic and molecular systems, the Hartree-Fock(HF) state is often an excellent choice. When low-lying excited states are of interest, reference states can be selected from low-order excited configurations. In condensed matter models, such as the Heisenberg or Hubbard models, approximate solutions from classical computations can serve as effective reference states for the quantum circuit. Following this preparation, a Hadamard gate (labeled as $H'$ in the diagram, with the matrix form $H' = \frac{\sqrt{2}}{2}\begin{pmatrix} 1 & 1 \\ 1 & -1 \end{pmatrix}$) is operated on the ancillary qubit $q_0$. The Hadamard gate transform the ancillary qubit into the following state:

$$|T_1\rangle = \frac{\sqrt{2}}{2}|0\rangle \otimes |\psi_{\text{ref}}\rangle + \frac{\sqrt{2}}{2}|1\rangle \otimes |\psi_{\text{ref}}\rangle \tag{1a}$$

Subsequently, the ancillary qubit $q_0$ controls a real-time Hamiltonian evolution operator $e^{-i\hat{H}t}$, which acts on the target qubit register. The operator $e^{-i\hat{H}t}$ is applied only when $q_0$ is in the $|1\rangle$ state, resulting in the following outcome:

$$|T_2\rangle = \frac{\sqrt{2}}{2}|0\rangle \otimes |\psi_{\text{ref}}\rangle + \frac{\sqrt{2}}{2}|1\rangle \otimes e^{-i\hat{H}t}|\psi_{\text{ref}}\rangle \tag{1b}$$

Finally, a Hadamard gate is applied to the ancillary qubit $q_0$ once again, resulting in the output state $|T_3\rangle$:

$$|T_3\rangle = \frac{1}{2}|0\rangle \otimes \left(I + e^{-i\hat{H}t}\right)|\psi_{\text{ref}}\rangle + \frac{1}{2}|1\rangle \otimes \left(I - e^{-i\hat{H}t}\right)|\psi_{\text{ref}}\rangle \tag{1c}$$

By measuring the ancillary qubit in the output state $|T_3\rangle$, the probabilities of collapsing to states $|0\rangle$ and $|1\rangle$ are determined as follows:

$$\begin{aligned} P_0 &= \left\|\frac{1}{2}\left(I + e^{-i\hat{H}t}\right)|\psi_{\text{ref}}\rangle\right\|^2 = \frac{1}{2} + \frac{1}{2}\langle\psi_{\text{ref}}|\cos(\hat{H}t)|\psi_{\text{ref}}\rangle \\ P_1 &= \left\|\frac{1}{2}\left(I - e^{-i\hat{H}t}\right)|\psi_{\text{ref}}\rangle\right\|^2 = \frac{1}{2} - \frac{1}{2}\langle\psi_{\text{ref}}|\cos(\hat{H}t)|\psi_{\text{ref}}\rangle \end{aligned} \tag{2}$$

By varying the evolution time $t$ within the quantum processor, these collapse probabilities can be measured at various intervals, resulting in the function $Q(t) = P_1(t) - P_0(t)$, which simplifies to:

$$Q(t) = \langle\psi_{\text{ref}}|\cos(\hat{H}t)|\psi_{\text{ref}}\rangle \tag{3}$$

Assuming the Hamiltonian $\hat{H}$ possesses eigenstates $|\Psi_0\rangle, |\Psi_1\rangle, \cdots, |\Psi_i\rangle, \cdots$ with corresponding eigenvalues $E_0, E_1, \cdots, E_i, \cdots$, any reference state $|\psi_{\text{ref}}\rangle$ can be expressed as a linear combination of these eigenstates, with coefficients $\{c_i\}$:

$$|\psi_{\text{ref}}\rangle = \sum_i c_i |\Psi_i\rangle \tag{4}$$

Substituting into $Q(t)$ and noting that $\hat{H}|\Psi_i\rangle = E_i|\Psi_i\rangle$, we have:

$$Q(t) = \sum_i c_i c_j^* \langle\Psi_j| \cos(\hat{H}t) |\Psi_i\rangle = \sum_i |c_i|^2 \cos(E_i t) \tag{5}$$

Assuming each eigenvalue $E_i$ is an integer ($K_i$) multiple of a minimal unit $\delta$, such that $E_i = K_i\delta \quad K_i \in Z$, and $\delta$ is sufficiently small to make the quantization error negligible, it follows:

$$Q(t) = \sum_i |c_i|^2 \cos(K_i \delta t) \tag{6}$$

Additionally, evaluating $Q(t + 2\pi/\delta)$ results in:

$$Q\left(t + \frac{2\pi}{\delta}\right) = \sum_i |c_i|^2 \cos\left(K_i\, \delta t + K_i\, \delta \cdot \frac{2\pi}{\delta}\right) = Q(t) \tag{7}$$

This demonstrates that $Q(t)$ is a periodic function with period $T = 2\pi/\delta$. When expanded into a Fourier series, $Q(t)$ is expressed as:

$$\begin{gathered} Q(t) = \frac{a_0}{2} + \sum_{n=1}^{\infty} [a_n \cos(n\omega t) + b_n \sin(n\omega t)] \\ a_n = \frac{1}{T}\int_{-\frac{T}{2}}^{\frac{T}{2}} Q(t)\cos(n\omega t)\, dt \qquad b_n = \frac{1}{T}\int_{-\frac{T}{2}}^{\frac{T}{2}} Q(t)\sin(n\omega t)\, dt \end{gathered} \tag{8}$$

where $\omega = 2\pi/T = \delta$, and given the even nature of $Q(t) = Q(-t)$, $b_n = 0$. It can be observed that Eq.6 and Eq.8 are formally identical. The Fourier coefficients $a_n$ directly correspond to $|c_i|^2$, and $n\omega$ to $K_i\delta$, essentially representing the eigenvalues $E_i$. Given the parity of the cosine function, $|n\omega| = |K_i\delta|$. Utilizing quantum computers, one can efficiently determine $Q(t)$ for any $t$. Subsequently, classical computers can be used to perform a Fourier transform to obtain the eigenvalues and calculate the overlaps between the eigenstates and the reference state.

Given that classical computers can only handle discrete signals and evaluating every time point within continuous intervals with quantum computers is impractical, it is necessary to define a reasonable sampling interval. This approach allows the measurement of $Q(t)$ at specific points, effectively converting a continuous signal into a format processable by classical computers. Sampling begins at $t = 0$, with an interval $\Delta$ and a total of $N$ samples: $t = 0, \Delta, 2\Delta \cdots N\Delta$. Quantum computers are utilized to capture $Q(t)$ values at each point, producing a series of discrete data points $q(n)$:

$$q(n) = Q(n\Delta) = \sum_i |c_i|^2 \cos(K_i\delta \cdot n\Delta) \tag{9}$$

The discretization of the time domain induces periodicity in the frequency domain, and vice versa. Let $R(k)$ represent the discrete Fourier transform of $q(n)$:

$$R(k) = \sum_{n=0}^{N-1} q(n) e^{-\boldsymbol{i}\frac{2\pi}{N}kn} \tag{10}$$

where $\boldsymbol{i}$ represents the imaginary unit. The corresponding inverse Fourier transform is:

$$q(n) = \frac{1}{N}\sum_{k=0}^{N-1} R(k) e^{\boldsymbol{i}\frac{2\pi}{N}nk} \tag{11}$$

The inverse transformation indicates that $q(n)$ is a periodic sequence with a period of $N$. Since $Q(t)$ satisfies $Q(t) = Q(-t)$, it follows that $q(n) = q(-n) = q(N-n)$. Substituting this into Eq. 10 demonstrates that $R(k)$ is an even function with a period of $N$. Based on the properties of the discrete Fourier transform, reconstructing the $q(n)$ sequence involves selecting $N$ consecutive $R(k)$ sequences starting from any point. For an odd number of samples, $N$, sample $R(k)$ from $-(N-1)/2$ to $(N-1)/2$:

$$q(n) = \frac{1}{N}\sum_{k=-\frac{N-1}{2}}^{\frac{N-1}{2}} R(k) e^{\boldsymbol{i}\frac{2\pi}{N}nk} = a_0 + \sum_{k=1}^{\frac{N-1}{2}} a_k \cdot \cos\left(\frac{2\pi}{N}nk\right) \tag{12}$$

Where $a_0 = R(0)/N \quad a_k = 2R(k)/N$, comparing Eq. 9 and Eq. 12, and considering the parity of the cosine function, we find:

$$\begin{gathered} |c_i|^2 = a_i \\ \left|2\pi \cdot \frac{k}{N} n\right| = |K_i\delta \cdot n\Delta| \end{gathered} \tag{13}$$

Eigenvalues are approximated as integer multiples of $\delta$, $E_i = K_i\delta$, with different $k$ values in $R(k)$ representing different multiples ($K_i$), thus indicating different eigenvalues. The corresponding equation is:

$$\frac{2\pi}{N} = \delta \cdot \Delta \qquad\qquad \frac{|k|}{N} = \frac{|E_i|}{2\pi} \cdot \Delta \tag{14}$$

Based on this analysis, the XZ24 algorithm for calculating eigenvalues is outlined as follows: First, select an appropriate reference state, such as the HF state for ground state energy calculations, or singly or doubly excited configurations for low-order excited states. Next, determine the sampling interval $\Delta$ and the total number of samples $N$. According to Eq. 14, the precision δ is defined by $N\Delta$, representing the maximum sampling evolution time $T_{\max} = N\Delta$. To achieve precise eigenvalues, δ needs to be as small as possible, necessitating long evolution time sampling of $Q(t)$. Once $N$ is set, due to the parity of $q(n)$, quantum computers need only sample from $n = 0$ to $n = N/2$, as the remaining values can be obtained from $q(n) = q(N - n)$. Perform the discrete Fourier transform on the $q(n)$ sequence to obtain its spectrum $R(k)$. According to Eq. 13, if $R(k)$ is non-zero, the corresponding frequency indicates an eigenstate with non-negligible overlap with the reference state, and its eigenvalues can be calculated using Eq. 13. It is worth noting that the quantum circuit sampling process introduces errors, and the quantization of eigenvalues $E_i$ is subject to inaccuracies. Therefore, when analyzing the $R(k)$ spectrum, emphasis should be on significant peaks, rather than all non-zero points.

## 3. Numerical simulation and discussion

### 3.1 Execution Process of the XZ24 Algorithm

To validate the effectiveness of the XZ24 algorithm, we tested the method using the $H_4$ molecular system and the Heisenberg model. For the $H_4$ molecular chain, whose structure is depicted in Fig. 2(a), calculations were performed using the STO-3G basis set. The detailed procedure is as follows: First, molecular orbitals were calculated using the PySCF software[59,60], and the Hamiltonian was represented in its second quantized form. Next, with the reference state set as the HF state, the circuits for $U_{\text{ref}}$ and $e^{-i\hat{H}t}$ were compiled into basic quantum gates using OpenFermion [62], thereby establishing the specific implementation of the quantum circuit shown in Fig. 1. It is noteworthy that $\hat{H}$ is a multi-qubit Hermitian operator, which can often be expressed as a sum of Pauli string products: $\hat{H} = \hat{h}_1 + \hat{h}_2 + \cdots + \hat{h}_Z$. To compile the operator $e^{-i\hat{H}t}$, the Suzuki–Trotter decomposition is generally employed [63–65]. The first- and second-order Trotter decompositions are given by:

$$\hat{U}_1(t) = e^{-i\hat{h}_1 t} e^{-i\hat{h}_2 t} \cdots e^{-i\hat{h}_Z t} \tag{15a}$$

$$\hat{U}_2(t) = e^{-\frac{i\hat{h}_1 t}{2}} e^{-\frac{i\hat{h}_2 t}{2}} \cdots e^{-i\hat{h}_Z t} \cdots e^{-\frac{i\hat{h}_2 t}{2}} e^{-\frac{i\hat{h}_1 t}{2}} \tag{15b}$$

In contrast to the first-order Trotter expansion, the second-order expansion provides improved accuracy and additionally preserves time-reversal symmetry: $\left[\hat{U}_2(\tau)\right]^\dagger = \hat{U}_2(-\tau)$ and $\hat{U}_2(\tau)\hat{U}_2(-\tau) = \hat{I}$.Given that the collapse probability of the ancillary qubit derived in Eq. (2) leverages the time-reversal characteristic of the operator $e^{-i\hat{H}t}$, we employ the second-order Trotter expansion to approximate the operator $e^{-i\hat{H}t}$ during circuit compilation.

Thirdly, an appropriate maximum sampling evolution time, $T_{\max}$, and sampling interval, $\Delta$, are set for the XZ24 algorithm. According to the preceding analysis, it is theoretically assumed that all eigen-energies can be expressed as integer multiples of a minimum unit $\delta$, with the calculation accuracy determined by this quantization unit $\delta$. To ensure that the results fall within chemical accuracy, Eq. (14) implies that the maximum sampling evolution time $T_{\max}$ should satisfy:

$$\delta = \frac{2\pi}{T_{\max}} < 0.0016 \tag{16}$$

Hence, $T_{\max} > 3927$. For convenience, we set $T_{\max} = 4000$. By selecting different sampling intervals $\Delta$, the total evolution time sampling numbers $N$ can be determined by $N = T_{\max}/\Delta$. Considering the parity and periodicity of $q(n)$, it is sufficient to sample from 0 to $N/2$. The remaining points can be obtained from $q(n) = q(N-n)$, effectively halving the total sampling. Fourthly, for each sampling point $t = n\Delta$, we substitute it into the quantum circuit shown in Fig. 1 and use a quantum processor to run the circuit. In this work, we use MindQuantum platform [66] to simulate the quantum circuits. The output state can be obtained, and the probability of the auxiliary qubit collapsing into the $|0\rangle$ or $|1\rangle$ states can be determined by sampling. Subtracting these probabilities yielded $Q(n\Delta)$, from which the values of $q(n)$ are derived. Finally, a discrete Fourier transform of the $q(n)$ sequence is performed on a classical computer. The eigenstate energies are determined from the spectral peaks of the resulting frequency spectrum.

### 3.2 Analysis of the Sampling Interval

To demonstrate that this method can effectively calculate both the equilibrium and dissociation states of molecular systems, Figs. 2(a) and 2(b) display the computational results for the $H_4$ molecular chain with bond lengths of 1.0 Å and 2.0 Å, respectively, which correspond to the spectra of $q(n)$. Figs. 2(c) to 2(f) present the $q(n)$ spectra for various Heisenberg models, where the Hamiltonians of these systems can all be expressed as:

$$\hat{H} = J\sum_{\langle j,k\rangle}\left[\sigma_j^x \cdot \sigma_k^x + \sigma_j^y \cdot \sigma_k^y\right] + h\sum_{\langle j,k\rangle}\sigma_j^z \cdot \sigma_k^z \tag{17}$$

Here, $J$ represents the spin exchange coupling in the X and Y directions at each lattice site, $h$ denotes the coupling in the Z direction, and $\langle j,k\rangle$ refers to the nearest-neighbor lattice sites. The operators $\sigma_j^x$, $\sigma_j^y$, and $\sigma_j^z$ correspond to the three Pauli matrices at site $j$. Figs. 2(c) and 2(d) show the computational results for a one-dimensional Heisenberg model with 10 lattice sites, while Figs. 2(e) and 2(f) present the $q(n)$ spectra of a two-dimensional 4×3 Heisenberg model. The corresponding models and reference states are depicted in Figs. 2(c) and 2(e). For the Hamiltonians corresponding to Figs. 2(c) and 2(e), we have $J = h = 1$ (XXX model), while for Figs. 2(d) and 2(f), the parameters are $J = 1 \quad h = 2$ (XXZ model).

The discrete Fourier transform spectra of the $q(n)$ signals of different quantum models with various sampling intervals are shown in Fig. 2. It is noteworthy that all $q(n)$ signals in Fig. 2 shares the same maximum sampling evolution time $T_{\max} = 4000$, hence different sampling intervals imply different total sampling numbers: $N = T_{\max}/\Delta$. As per the previous analysis, peak points in the spectra represent various eigenstates of the Hamiltonian. The x-coordinates denote different frequencies: $x_k = \frac{k}{N\Delta}$, while the y-coordinates represent the Fourier transform coefficients of different frequencies $a_k$. According to Eq. 14, $|E_i| = 2\pi x_k$. For the convenience of comparison, some eigenstates with non-negligible overlaps with the reference state are displayed using dotted vertical lines. These eigenstates are obtained via the exact diagonalization (ED) method. Theoretically, the positions of these lines will coincide with peaks in the spectrum.

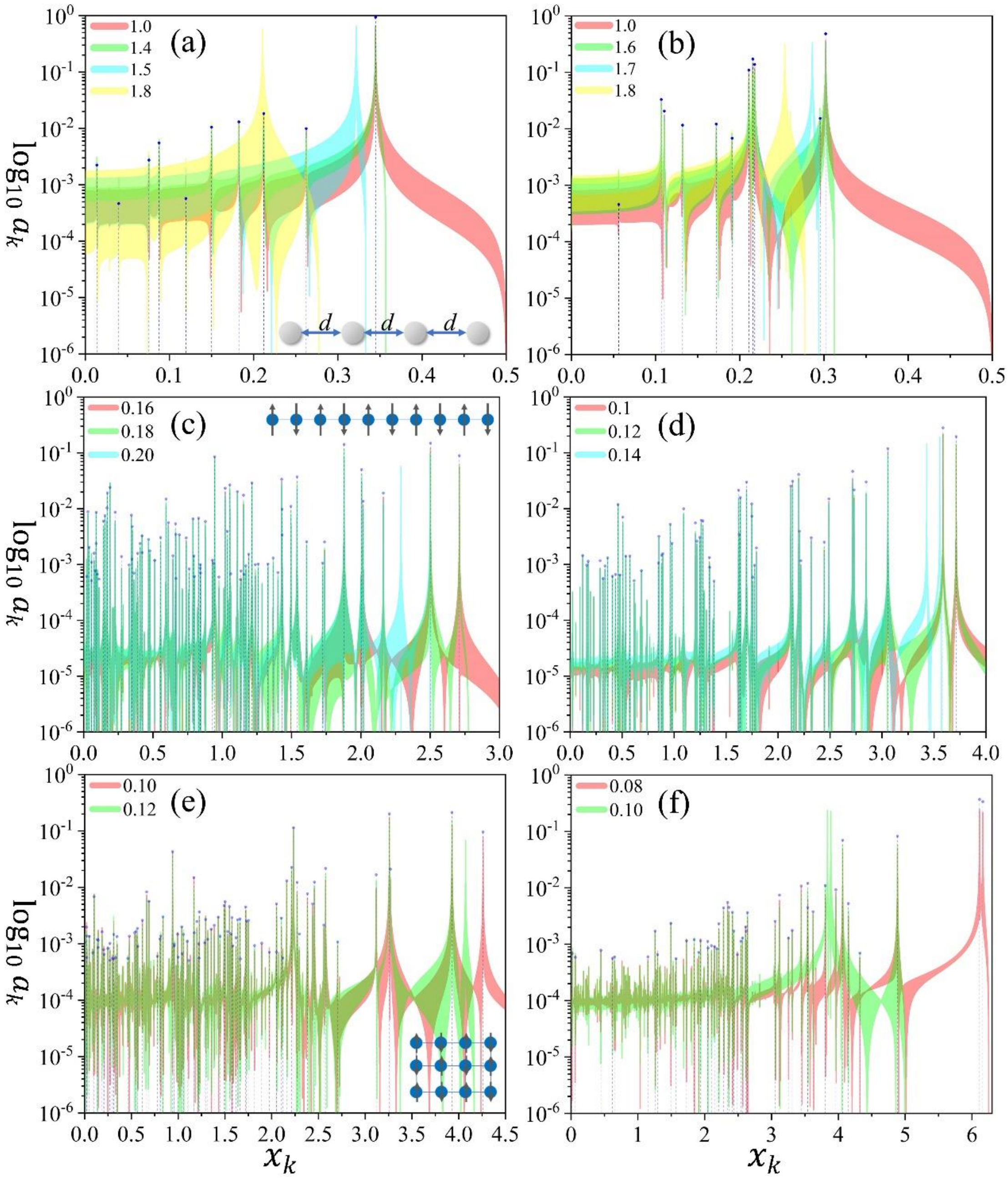

Figure 2. Discrete Fourier transform spectra of the $q(n)$ signals for various quantum systems: $H_4$ molecular chain with bond lengths of 1.0Å(a) and 2.0Å(b), 10-sites of one-dimensional Heisenberg model with $J = h = 1$(c) and $J = 1 \quad h = 2$(d), 4×3 two-dimensional Heisenberg model with $J = h = 1$(e) and $J = 1 \quad h = 2$(f). The computational models are illustrated in (a), (c), and (e), with arrows at each lattice point in models (c) and (e) representing the Z-spin components, which are used to establish the reference state. Conversely, (a) and (b) employ the HF state as their reference state. The sampling time intervals ($\Delta$) are indicated in the legend. The x-axis is defined as $x_k = k/N\Delta$, and the y-axis represents $\log_{10} a_k$. The blue dotted vertical lines indicate the eigenstates obtained from the exact diagonalization method, where only eigenstates with overlaps greater than $5 \times 10^{-4}$ are displayed, i.e., eigenstates satisfying $|c_i|^2 \geq 5 \times 10^{-4}$ in Eq. 4. The x-coordinates of these lines are $|E_i|/2\pi$, and their heights are $|c_i|^2$.

Fig. 2 demonstrates that when the sampling interval is below the threshold, although the shapes of the spectrum curves may vary, the peaks of all curves align perfectly with the vertical lines representing the eigenvalue (ED) positions. However, when the sampling interval exceeds the threshold, the peaks no longer align with the ED lines. The threshold is system-dependent. For an $H_4$ molecular chain with a bond length of 1.0 Å, the sampling interval must not exceed 1.4. When the bond length increases to 2.0 Å, the sampling interval must be less than 1.6. In the case of a one-dimensional Heisenberg model with 10 sites, the sampling interval should not exceed 0.18 when $J = h = 1$. For $J = 1 \quad h = 2$, even a sampling interval of 0.14 would cause deviations in the peak positions of the spectrum. For a 4×3 two-dimensional Heisenberg model, the threshold is even smaller. A sampling interval of 0.1 is suitable for $J = h = 1$, but not for the $J = 1 \quad h = 2$ model. Additionally, Fig. 2 demonstrates that even if the sampling interval exceeds this threshold, not all spectral peaks exhibit deviations. Typically, peaks corresponding to eigen-energies with larger absolute values shift, while those corresponding to eigen-energies with smaller absolute values remain accurate.

The primary reason for this phenomenon lies in the excessively long sampling intervals. According to the Nyquist-Shannon sampling theorem, if a signal's maximum frequency is $f_{\max}$, in order to reconstruct the original signal without distortion, the sampling frequency must be at least $2f_{\max}$. For the $Q(t)$ signal, based on Eq. 5, its frequencies are determined by various of eigen-energies $E_i$: $f_i = \frac{|E_i|}{2\pi}$. Therefore, the maximum frequency of $Q(t)$ is determined by the eigen-energy with the largest absolute value. Generally, for any quantum system, the ground state energy often has the largest absolute value: $f_{\max} = \frac{|E_0|}{2\pi}$. Given that the ground state energy of the $H_4$ system with a bond length of 1.0 Å is -2.1664 Ha, the maximum sampling interval can be determined as:

$$\Delta_{\max} = \frac{1}{2f_{\max}} = \frac{\pi}{|E_0|} = 1.45 \tag{18}$$

Thus, for spectra with $\Delta = 1.5$ or 1.8, as shown in Fig. 2(a), the sampled $q(n)$ signal experiences spectral aliasing, leading to an inability to accurately reconstruct the original spectral graph. For the systems shown in Fig. 2(b)–(f), the maximum sampling intervals are 1.66, 0.184, 0.135, 0.117, and 0.081, which align with the observations in the figure. To reduce sampling cost, the algorithm can utilize the largest possible sampling interval during execution. Although the exact eigenenergy with the largest absolute value (often can be seen as the ground state energy) is unknown, classical approximation algorithms can be employed to estimate an upper bound of the ground state energy $E_0$. By substituting this upper bound into Eq. (18), a relatively optimal value for $\Delta$ can be determined. Given the approximate proportionality between the magnitude of the ground state energy and the system size $L$, such as in the case of an infinite one-dimensional Heisenberg model where the eigen-energies asymptotically scale linearly with the number of lattice sites, we find that: $\Delta \propto 1/|E_0| \propto 1/L$.

It is important to note that, in theory, based on Eq. (13), the height of a spectral peak directly reflects the overlap between the reference state and the corresponding eigenstate, i.e., $|c_i|^2 = a_i$. However, even when the sampling interval satisfies the Nyquist sampling theorem, spectral leakage can still occur due to non-periodic sampling, resulting in $|c_i|^2 \neq a_i$. While this causes quantitative inaccuracies in calculating the wavefunction overlap, the qualitative analysis remains reliable. In other words, the height of each spectral peak can still be reliably used to identify which eigenstates dominate in the reference state.

### 3.3 Determining the Sign of Eigenvalues

It is worth noting an issue of the proposed algorithm: XZ24 can only determine the absolute value of the eigen-energy, but not its sign. However, this is not a significant problem. Firstly, for the majority of quantum systems, eigen-energies are typically negative, so when calculating some low-energy eigenvalues, their absolute values can be directly taken as negative. Secondly, even if the sign cannot be determined, it can be inferred by applying a small positive constant offset $s_0$ to the Hamiltonian: the eigenvalues of the original Hamiltonian $\widehat{H}$ are $E_i$,and those after the offset are $E_i + s_0$. When $E_i$ is positive, $|E_i + s_0|$ will be greater than $|E_i|$, resulting in a right shift of the spectral peak corresponding to $E_i$; when $E_i$ is negative, $|E_i + s_0|$ will be smaller than $|E_i|$, resulting in a left shift of the peak in the corresponding spectrum.

Therefore, by performing the XZ24 algorithm separately on the Hamiltonian $\widehat{H}$ and $\widehat{H} + s_0$, and comparing their spectra, the direction of the peak shift can determine the sign of the eigenvalue. As shown in Fig. 3, for a 3×3 two-dimensional Heisenberg model with $J = h = 1$, the leftward shift of the peaks near 1.34, 2.4, and 3.0 along the x-axis indicates negative eigen-energies, whereas the rightward shift of the peaks near 0.95 and 1.9 suggest positive eigen-energies, both consistent with the ED results.

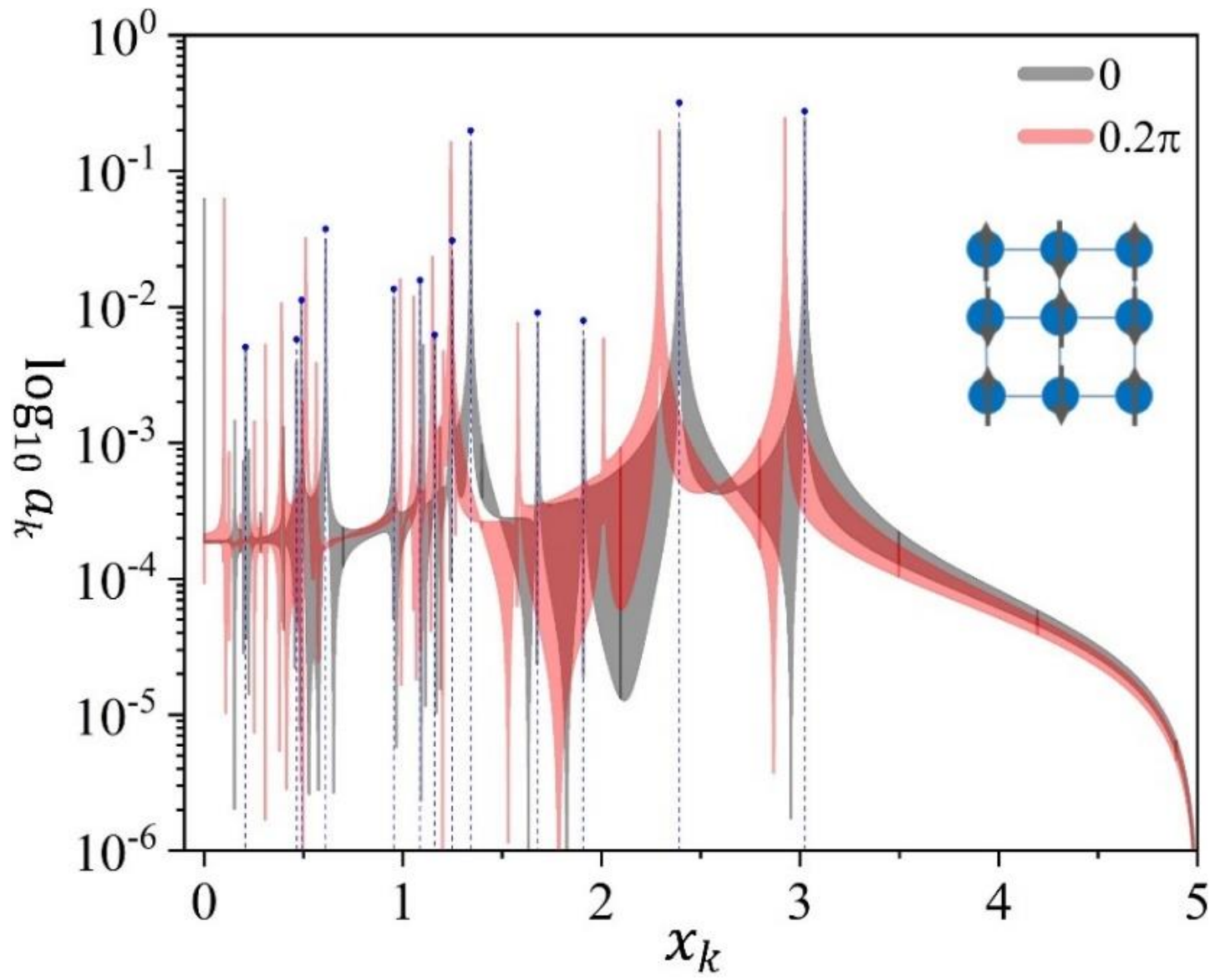

Figure 3: Discrete Fourier transform spectra of the $q(n)$ signals for a 3×3 two-dimensional Heisenberg model with $J = h = 1$, where $T_{\max} = 1000$ and $\Delta = 0.1$ for all sampling sequences. The x-axis, y-axis, and blue dotted vertical lines have the same meaning as those in Figure 2. The only distinction between the black and red lines is that the red line has a positive constant offset of $0.2\pi$ applied to the Hamiltonian, while the black line has no offset.

### 3.4 Analysis of Computational Accuracy

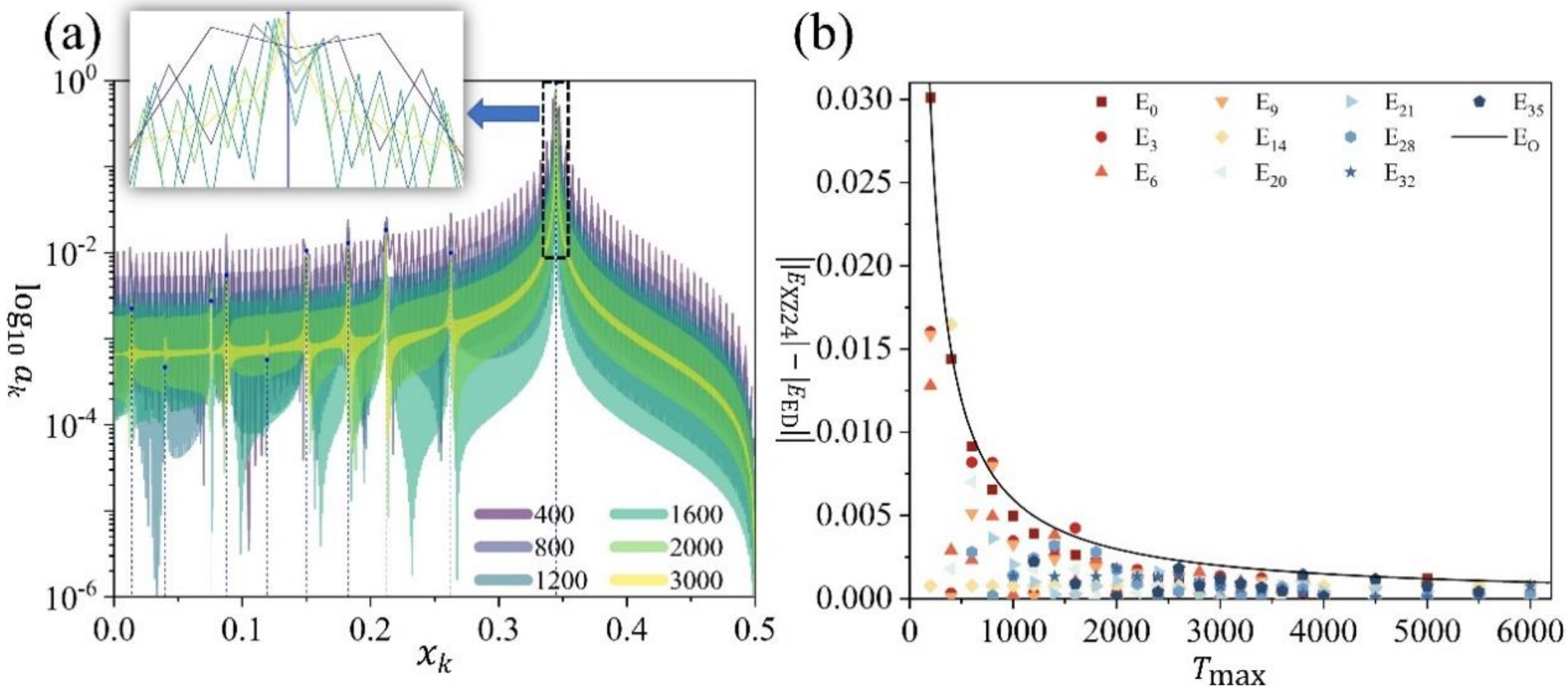

Figure 4: The XZ24 calculation results for an $H_4$ molecular chain with a bond length of 1.0 Å. Panel (a) shows the spectra of $q(n)$ signals with $\Delta = 1.0$, but for different maximum evolution times ($T_{max}$), as indicated in the legend. The x-axis, y-axis, and blue dotted vertical lines have same meanings with those in Figure 2. The figure also includes a magnified view of the spectral plot, focusing on the x-axis range between 0.34 and 0.35. Panel (b) depicts the computational errors for the $q(n)$ signals, again with $\Delta = 1.0$ but with varying $T_{max}$. The y-axis represents the absolute difference between the eigen-energies computed by the XZ24 method and those obtained via ED for each eigenstate. The x-axis shows different values of $T_{max}$, with points corresponding to various eigenstates. The black solid line represents the fitted maximum error as a function of $T_{max}$.

The theoretical analysis indicates that the computational error is solely determined by the maximum evolution time. This is demonstrated numerically using an $H_4$ molecular chain with a bond length of 1.0 Å, as shown in Fig. 4. In most cases, the eigen-energies $E_i$ cannot be expressed as integer multiples of $\delta$, leading to quantization errors. To minimize these errors, the quantization unit $\delta$ should be as small as possible. From the analysis, $\delta = 2\pi/(N \cdot \Delta) = 2\pi/T_{max}$, implying that the longer the maximum evolution time, the higher the spectral resolution. Fig. 4(a) presents spectra of various $q(n)$ sequences obtained with $\Delta = 1.0$ under different $T_{max}$, with ED values serving as the benchmark. As shown, none of the selected $T_{max}$ values allow the ground state energy to be an exact integer multiple of $\delta$. Consequently, all spectral peaks around $x_k = 0.35$ deviate from the ground state energy. However, as $T_{max}$ increases, the deviation significantly decreases due to improved spectral resolution. Fig. 4(b) quantitatively describes the relationship between this deviation and $T_{\max}$ by showing the absolute deviation between the eigenvalues from the spectral peak positions and those obtained via ED. Since the XZ24 method can only resolve eigenstates with non-zero overlap with the reference state, only the errors for computable eigenvalues are shown. For some eigen-energies (e.g., $E_{14}$), the deviation remains small even when $T_{\max}$ is short (i.e., $\delta$ is large), primarily due to these eigen-energies being close to integer multiples of $\delta$, a result of fortuitous coincidence. Nevertheless, even when eigenenergies deviate from integer multiples of $\delta$, the deviation will not exceed $\delta$, making $\delta$ the upper bound for the error. From Fig. 4(b), it is evident that at a constant sampling frequency, the upper limit of computational error shows an approximately inverse relationship with $T_{\max}$. To elucidate this, we fit the maximum error with the function $y = A/x$, yielding $E_{\text{err}} = 5.964/T_{\max}$. The fit exhibits a high degree of accuracy with $R^2 = 0.98695$, and the fitted coefficient of 5.964 is very close to $2\pi$, further validating the correctness of the theoretical analysis presented.

Unlike the requirement for chemical accuracy in quantum chemistry calculations, many quantum models prioritize relative accuracy, expressed as $|E_{XZ24} - E_{ED}|/|E_{ED}|$. Based on the analysis above, the absolute error $|E_{XZ24} - E_{ED}|$ is determined by $\delta = \frac{2\pi}{T_{\max}} \geq |E_{XZ24} - E_{ED}|$, while the maximum sampling interval is determined by the ground state energy: $\Delta_{\max} = \pi/|E_0|$. Therefore, according to Eq. (14), when $\Delta$ is set to its maximum value, the total number of sampling steps for the evolution time is:

$$N = \frac{2\pi}{\Delta_{\max}\delta} \leq \frac{2|E_0|}{|E_{XZ24} - E_{ED}|} \tag{19}$$

This implies that once the maximum relative error is determined, $N$ is also fixed. According to Eq. (19), when the relative error is set to 0.1%, sampling only 2000 evolution time points, with the sampling interval at its maximum value, ensures that the calculated ground state energy meets the relative error constraint. However, since this method is also used to compute higher-energy eigenstates, and given that the absolute error remains constant while the eigen-energies $|E_i|$ for higher states are smaller, the resulting relative error will be larger for those states. Additionally, because the exact value of the ground state energy is unknown in advance, the sampling interval $\Delta$ must be set slightly smaller than $\Delta_{\max}$ during the computation process. These factors all affect the final relative error. Therefore, when executing the algorithm, some redundancy in the sampling number $N$ is necessary; for instance, setting $N = 4000$ ensures that the relative error remains below 0.1%. (Of course, considering the symmetry of the $q(n)$ sequence, where $q(n) = q(N - n)$, only the sequence from $n = 0$ to $n = 2000$ needs to be sampled on the quantum processor.) To numerically demonstrate this, Fig. 5 presents the relative error obtained using the XZ24 method for various Heisenberg models. In all cases, the sampling number $N$ is set to 4000, and the sampling interval $\Delta$ is slightly smaller than their respective $\Delta_{\max}$.

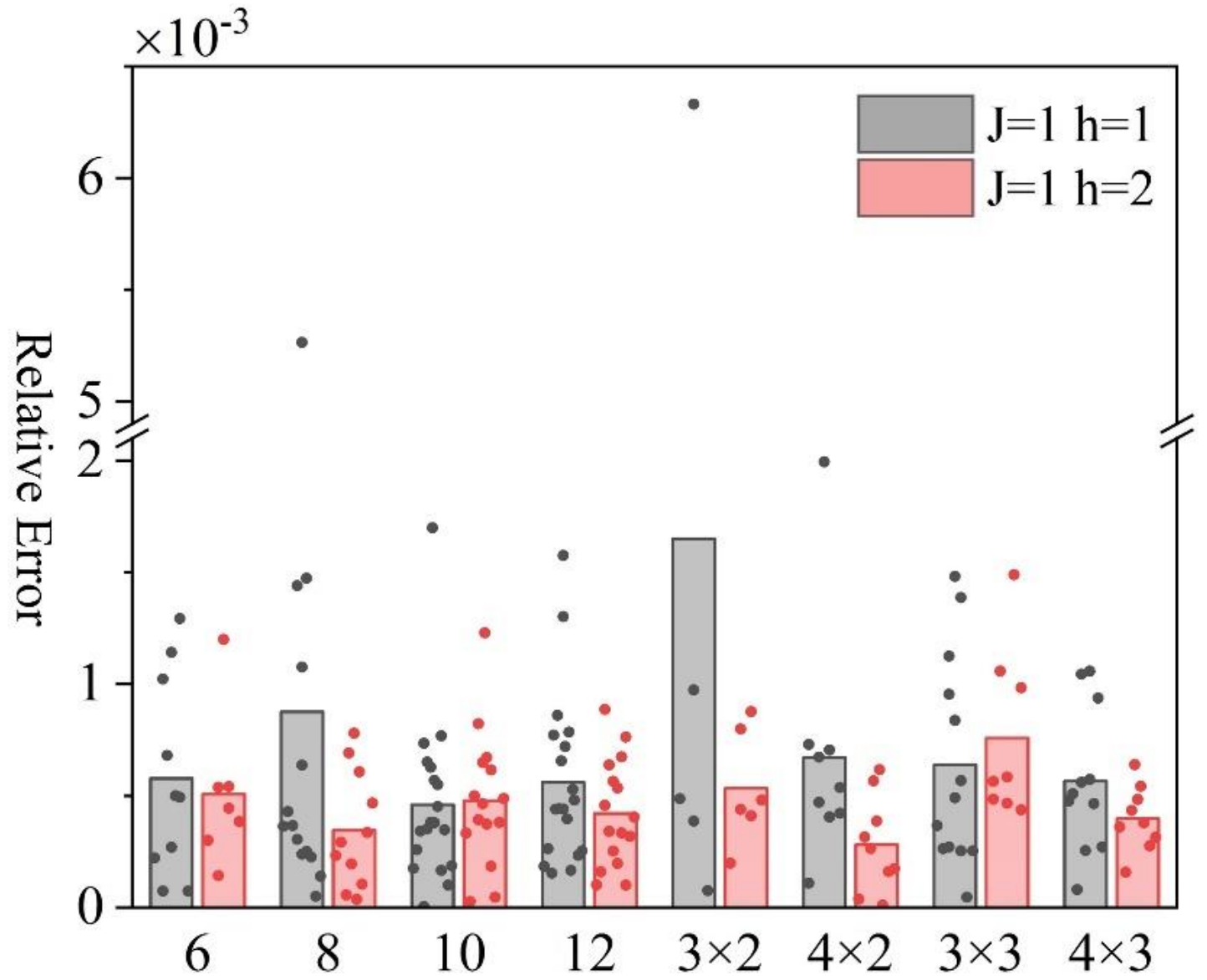

Figure 5: Relative errors in the eigen-energies of various Heisenberg models computed using the XZ24 method. The X-axis represents different models, ranging from left to right: one-dimensional Heisenberg models with 6, 8, 10, and 12 lattice sites, followed by two-dimensional Heisenberg models with lattice configurations of 3×2, 4×2, 3×3, and 4×3. The Y-axis represents the relative error, $|E_{XZ24} - E_{ED}|/|E_{ED}|$, where $E_{ED}$ denotes the eigenenergies calculated using the ED method. Red and black colors represent different model parameters, as specified in the legend. For all calculations, the sampling number $N$ is set to 4000, and the sampling interval $\Delta$ are set as follows, from left to right: 0.3, 0.2, 0.2, 0.15, 0.18, 0.12, 0.15, 0.1, 0.25, 0.18, 0.16, 0.11, 0.18, 0.13, 0.1, and 0.08. Each point in the figure represents the relative error for one eigenvalue of the respective model, and the bars represent the average relative error across all eigenvalues for that model.

In Fig. 5, the relative error for most of the calculated results remains within 0.1%, and the average relative error across all models also stays below 0.1%, with the exception of the 3×2 two-dimensional Heisenberg model with parameters $J = h = 1$.

The larger average relative error in this model is due to one eigenstate with an energy of 0.05059, while the XZ24 method computes it as 0.05027. Although the absolute error between these values is small, the eigenvalue itself is close to zero, resulting in an inflated relative error. Similar reasons explain the relatively larger errors in other eigenvalues shown in Fig. 5. In general, we are primarily concerned with low-energy eigenstates, whose eigen-energies are not far from the ground state and tend to have larger absolute values. Therefore, the likelihood of excessively large relative errors occurring for these eigenstates is low.

### 3.5 Analysis of the Quantum State Measurement Count

All the preceding analysis and discussion are predicated on the assumption that the $q(n)$ sequence can be sampled with high accuracy. In the implementation of the XZ24 algorithm, obtaining $q(n)$ requires measuring the probability of the ancillary qubit collapsing into the $|1\rangle$ and $|0\rangle$ states, with the difference between these probabilities providing the value of $q(n)$. However, due to the finite number of quantum state measurements, accurately determining the collapse probabilities is challenging, and the resulting $q(n)$ values are inherently subject to sampling errors. To evaluate the impact of these errors on the algorithm's performance, Fig. 6 presents the effect of different quantum state measurement counts $M$ on the $q(n)$ spectrum for various systems. Comparing the spectra for various measurement counts reveals that sampling error appears as white noise superimposed on the error-free $q(n)$ spectrum. Unlike the error-free $q(n)$ signal, this white noise is uniformly distributed across all frequency components, with random amplitudes and a consistent power density across the spectrum, ensuring that no single frequency is dominant. Consequently, when the number of quantum state measurements is low, eigenstates with small $a_k$ values in the $q(n)$ spectrum may be obscured by the white noise, rendering them difficult to discern. Nevertheless, the accuracy of eigenvalue estimates for more prominent peaks (those with larger $a_k$ values) remains unaffected. As the number of measurements $M$ increases, the overall amplitude of the white noise decreases, allowing lower peaks to become distinguishable, and the associated eigenvalues can be obtained via Eq. (14).

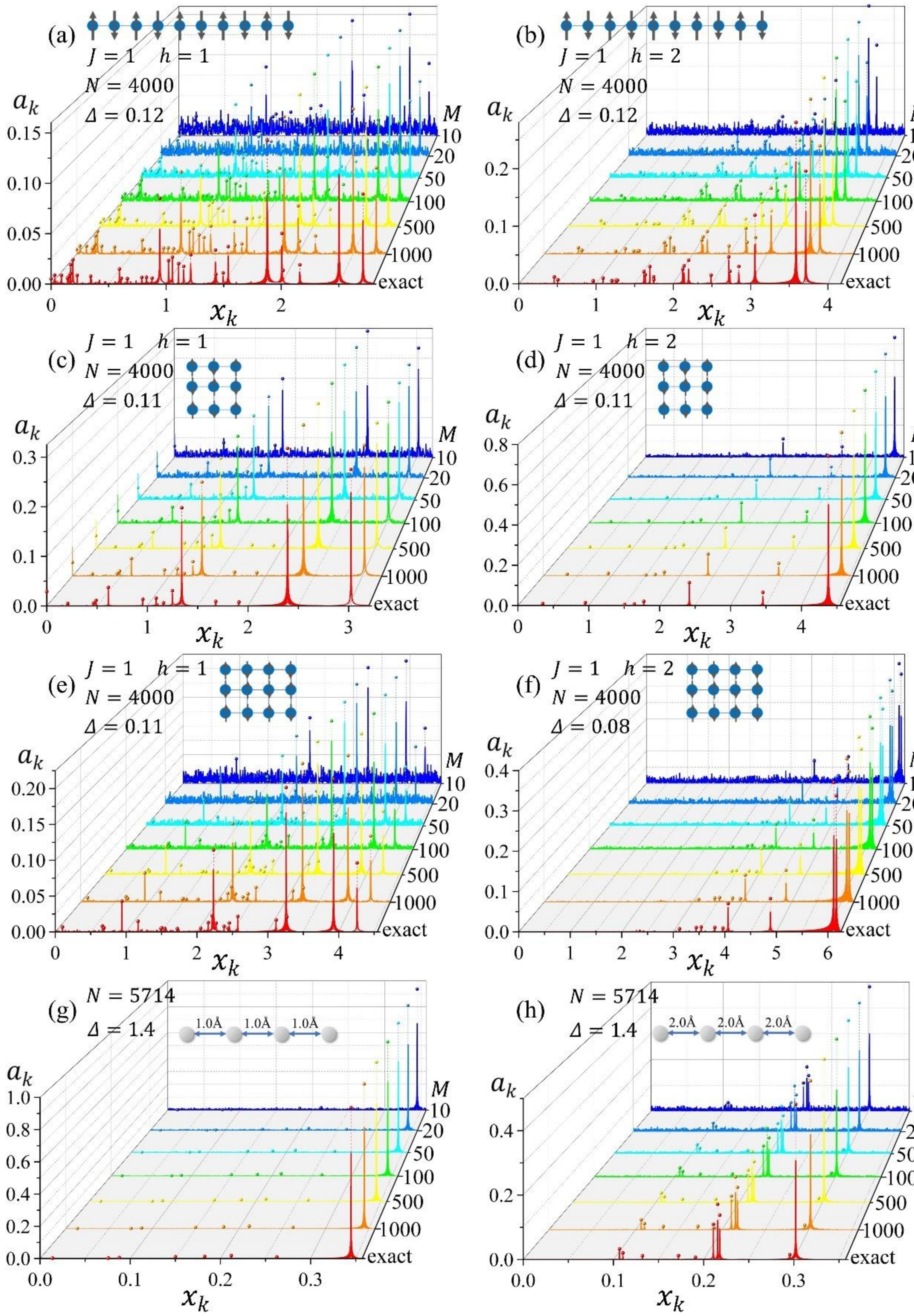

Figure 6: The spectral plots of $q(n)$ sequences estimated under different quantum circuit sampling numbers (M=10,20,50,100,200,500,1000, and 'exact') for various Heisenberg models and the $H_4$ molecular chain. The configurations of the different models, the reference states selected for the calculations, and other relevant computational parameters are annotated in the figure. The $x_k$, $a_k$ and the dotted vertical lines retain the same meanings as in Fig. 2. The different colored spectral lines represent different values of $M$, with 'exact' corresponding to an infinite circuit sampling number, indicating the results without sampling errors.

As shown in Fig. 6(a), when $M = 10$, the small number of samples results in significant white noise interference with high amplitude in the spectrum. However, distinct peaks at 0.944,1.877,2.502 and 2.71 in X-axis still stand out clearly from the white noise, allowing the corresponding eigenstates to be identified. Although there is also a peak at $x_k = 2.006$, its height is close to the white noise amplitude, making it difficult to confidently determine whether it represents a genuine eigenvalue or a random peak caused by sampling error. Other peaks with smaller $a_k$ values are completely buried in the white noise from sampling error. When $M = 50$, the peak at $x_k = 2.006$ shows a clear distinction from the white noise, enabling identification of the corresponding eigenstate. Additional peaks appear at 0.190, 1.021, 1.427, and 1.540 in $x_k$ positions, which can also be confirmed. At $M = 1000$, the white noise amplitude becomes very small, and most peaks corresponding to eigenstates become visible. Furthermore, all peak positions match exactly with the peaks in the error-free $q(n)$ spectrum. Similar phenomena are observed in calculations for other systems. In Fig. 6(c), only three eigenvalues can be identified when $M = 10$, six eigenvalues at $M = 100$, and eleven eigenvalues at $M = 1000$. In Fig. 6(g), since the reference state has a near-perfect overlap with the ground state but much smaller overlap with other eigenstates, the ground state is easily identified, but detecting other eigenstates requires significantly increasing the number of measurements. When $M = 10$, only the ground state is found; at $M = 100$, three eigenvalues can be identified; and at $M = 1000$, six eigenvalues are detected.

In conclusion, while the number of quantum state measurements does not impact the accuracy of the computational results, sampling errors caused by a finite number of measurements introduce white noise across the entire frequency spectrum. This noise, with random amplitudes at individual frequency points but uniformly distributed across the band, can obscure eigenstates with smaller peak values, thereby reducing the algorithm's efficiency in identifying eigenstates. When the overlap between the chosen reference state and an eigenstate's wavefunction is large, fewer quantum measurements are sufficient to calculate the corresponding eigenenergy. Conversely, when the overlap is small, increasing the number of measurements becomes necessary.

The preceding discussion qualitatively outlines the influence of the quantum measurement count on the performance of the XZ24 algorithm. To further this analysis, we now introduce a quantitative framework by developing a statistical model that accounts for sampling errors and explores the spectral distribution of these errors. Specifically, we aim to determine the upper bound of the peak height in the noise spectrum resulting from sampling errors when the sampling number of the evolution time is $N$, and the number of quantum measurements is $M$. If the observed peak exceeds this theoretical upper bound, it can be inferred that the peak is associated with an eigenstate, thus providing a theoretical basis for identifying spectral peaks. In the following derivation, we denote the sampling error as $s(n)$, and the estimate of $q(n)$, which incorporates the sampling error, as $\bar{q}(n)$. This relationship can be expressed as follows:

$$\bar{q}(n) = q(n) + s(n) \tag{20}$$

For a qubit, the probabilities of collapsing to the $|0\rangle$ and $|1\rangle$ states are denoted as $P_0$ and $P_1$, respectively. Through calculation, the expected value of $\bar{q}(n)$ for a single measurement is $E[\bar{q}(n)] = q(n) = P_1 - P_0$, and its variance is $D[\bar{q}(n)] = 4P_0P_1$. This indicates that the sampling variance is inherently linked to the state of the qubit. Since $0 \leq 4P_0P_1 \leq 1$, we simplify the subsequent analysis by assuming the worst-case scenario, i.e., $D[\bar{q}(n)] = 1$. According to the law of large numbers, when the quantum state is measured $M$ times, $\bar{q}(n)$ follows a normal distribution with an expectation of $q(n)$ and a variance of $1/M$. Substituting this result into Eq. (20), we find that the random signal $s(n)$ follows a normal distribution $N(0,1/M)$. To reduce sampling costs, the symmetry $\bar{q}(n) = \bar{q}(-n)$ is utilized when sampling $\bar{q}(n)$ sequences at different $n$ values during algorithm execution. This symmetry implies that $s(n)$ also satisfies $s(n) = s(-n)$. By analogy with Eq. (12), the forward and inverse Fourier transforms of $s(n)$ are given by:

$$\begin{aligned} S(k) &= \sum_{n=0}^{N-1} s(n) e^{-i\frac{2\pi}{N}kn} = s(0) + 2\sum_{n=1}^{\frac{N-1}{2}} s(n)\cos\left(\frac{2\pi}{N}nk\right) \\ s(n) &= \frac{1}{N}\sum_{k=0}^{N-1} S(k) e^{i\frac{2\pi}{N}nk} = b_0 + \sum_{k=1}^{\frac{N-1}{2}} b_k \cos\left(\frac{2\pi}{N}nk\right) \end{aligned} \tag{21}$$

where $b_0 = S(0)/N$ and $b_k = 2S(k)/N$. It can be observed that $S(k)$ is a linear combination of the $N$-term sequence $s(n)$. Since each $s(n)$ is independent and follows the $N(0,1/M)$ distribution, by the properties of the normal distribution, $S(k)$ also follows a normal distribution, with the expectation and variance given by:

$$\begin{gathered} E[S(k)] = 0 \\ D[S(k)] = \frac{1}{M} + \frac{1}{M} \cdot 4 \sum_{n=1}^{\frac{N-1}{2}} \cos^2\left(\frac{2\pi}{N} nk\right) \end{gathered} \tag{22}$$

For the term $\sum_{n=1}^{(N-1)/2} \cos^2\left(\frac{2\pi}{N} nk\right)$, when $k = 1$, we have:

$$\begin{aligned} \sum_{n=1}^{\frac{N-1}{2}} \cos^2\left(\frac{2\pi}{N} n\right) &= \sum_{n=1}^{\frac{N-1}{4}} \cos^2\left(\frac{2\pi}{N} n\right) + \sum_{n=\frac{N+3}{4}}^{\frac{N-1}{2}} \cos^2\left(\frac{2\pi}{N} n\right) \\ &= \sum_{n=1}^{\frac{N-1}{4}} \left[\cos^2\left(\frac{2\pi}{N} n\right) + \cos^2\left(\frac{2\pi}{N}\left(n + \frac{N-1}{4}\right)\right)\right] \end{aligned} \tag{23}$$

Using $\cos\left(\theta + \frac{\pi}{2}\right) = -\sin(\theta)$, we obtain:

$$\cos\left(\frac{2\pi}{N}(n + \frac{N-1}{4})\right) = -\sin\left(\frac{2\pi}{N} n - \frac{\pi}{2N}\right) \tag{24}$$

When $N$ is large, $\frac{\pi}{2N} \to 0$, and we can approximate the $\cos^2\left(\frac{2\pi}{N}\left(n + \frac{N-1}{4}\right)\right)$ term using the first-order Taylor expansion:

$$\begin{aligned} \cos^2\left(\frac{2\pi}{N}\left(n + \frac{N-1}{4}\right)\right) &\approx \left[\sin\left(\frac{2\pi}{N} n\right) - \frac{\pi}{2N} \cdot \cos\left(\frac{2\pi}{N} n\right)\right]^2 \\ &\approx \sin^2\left(\frac{2\pi}{N} n\right) - \frac{\pi}{2N} \cdot \sin\left(\frac{4\pi}{N} n\right) \end{aligned} \tag{25}$$

Substituting this into Eq. (23), we get:

$$\begin{aligned} \sum_{n=1}^{\frac{N-1}{2}} \cos^2\left(\frac{2\pi}{N} n\right) &= \sum_{n=1}^{\frac{N-1}{4}} \left[\cos^2\left(\frac{2\pi}{N} n\right) + \sin^2\left(\frac{2\pi}{N} n\right) - \frac{\pi}{2N} \cdot \sin\left(\frac{4\pi}{N} n\right)\right] \\ &= \frac{N-1}{4} - \frac{\pi}{2N} \cdot \sum_{n=1}^{\frac{N-1}{4}} \sin\left(\frac{4\pi}{N} n\right) = \frac{N-1}{4} \end{aligned} \tag{26}$$

The above derivation relies on the conclusion that the $4\pi n/N$ represents angles that are uniformly distributed on the unit circle. As $N$ is large, the sine values of these angles tend to cancel each other out. Substituting this into Eq. (22), we have $D[S(1)] = N/M$. Furthermore, for any frequency $k$, we can similarly derive the variance of $S(k)$, expressed as follows:

$$D[S(k)] = \frac{N}{M} \tag{27}$$

And given that $b_k = 2S(k)/N$, we get:

$$E[b_k] = 0 \qquad D[b_k] = \frac{4}{NM} \tag{28}$$

Considering that the amplitude at each frequency in the spectrum is taken as the absolute value of the Fourier transformed result, the signal $s(n)$ will be represented as $|b_k|$ form in the frequency-amplitude spectrum. Therefore, the sampling error will be superimposed onto the spectrum of $q(n)$ in the form of a half-normal distribution: $|b_k| \sim \text{Half Normal}\left(0, \frac{8}{NM}\right)$. The scale parameter for $|b_k|$, which corresponds to the standard deviation of the normal distribution, is $\sigma = \sqrt{\frac{8}{NM}}$. Based on the properties of the half-normal distribution, we know that the probability of $|b_k|$ falling within $3\sigma$ is 0.999978. This implies that if a spectral peak with an amplitude greater than $3\sigma$ is observed, the probability that it originates from the random noise caused by sampling error is only 0.00002 (i.e., 1 in 50,000). Thus, it can be concluded with high confidence that such a peak originates from the spectrum of $q(n)$ without sampling error, i.e. from the eigenstate.

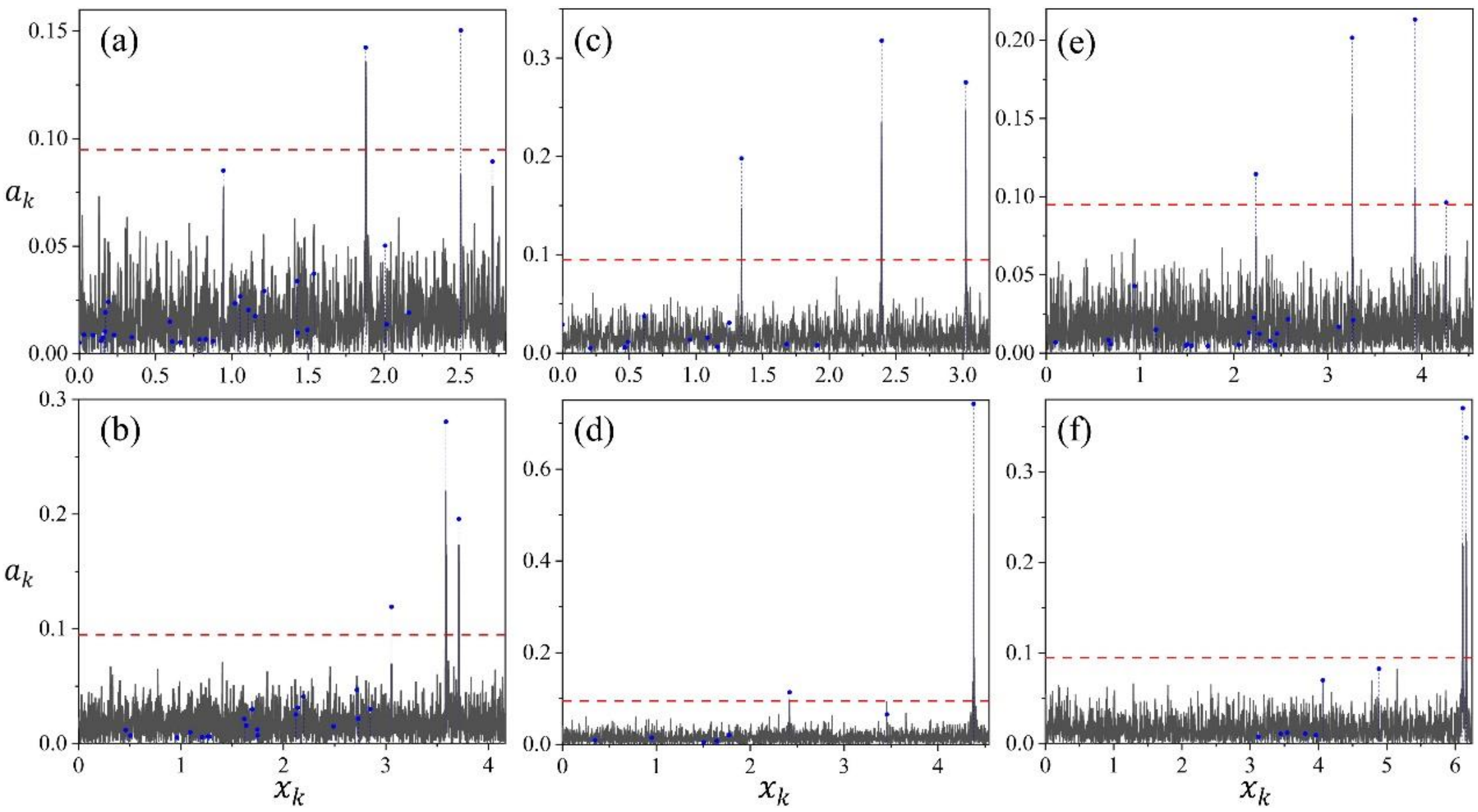

Figures 7(a)~(f) correspond to the spectra of $q(n)$ shown in Figures 6(a)~(f), where the number of quantum measurements is set to $M = 2$. The blue dotted vertical lines have the same meaning as those in Figure 2. The horizontal red dashed line represents the threshold height of $3\sigma = 0.0949$.

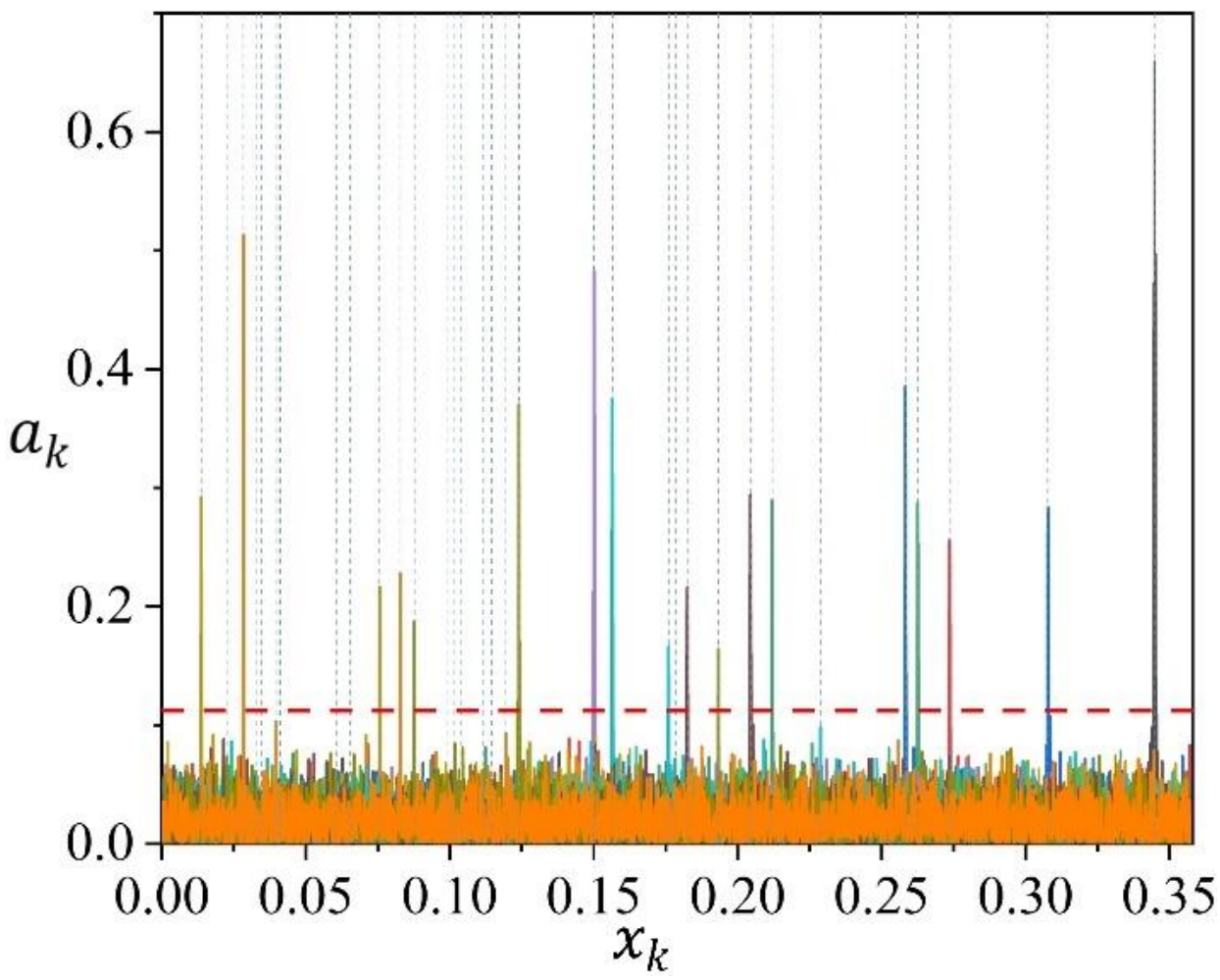

Figure 8 corresponds to the $q(n)$ spectrum in Fig. 6(g), where the number of quantum measurements is set to 1. The curves in different colors represent the $q(n)$ spectrum calculated under different reference states. The blue vertical dashed lines indicate the positions of all eigenstates, while the red horizontal dashed line represents the threshold height of $3\sigma = 0.1123$.

The above analysis provides a reliable criterion for peak identification: based on the number of evolution time samples, $N$, and the number of quantum state measurements, $M$, a threshold parameter $3\sigma$ can be established to filter the spectral peaks. Only peaks with heights exceeding this threshold are accepted, and the corresponding eigenstates can be computed using Eq. (14). As demonstrated, when $N$ is sufficiently large, if there is significant overlap between the eigenstate to be determined and the reference state, sampling noise will barely obscure these eigenstates, even when the quantum state measurement count $M$ is small. For instance, in the calculations shown in Fig. 6(a)~(f), where the evolution time sample number is $N = 4000$, if we are only interested in eigenstates with overlap $|c_i|^2 > 0.1$, measuring the quantum state just twice (i.e., $M = 2$, where the possible values of $\bar{q}(n)$ are limited to -1, 0, and 1) is sufficient, as the threshold $3\sigma = 0.095$ under this condition. To verify our analysis, Fig. 7(a)~(f) replicate the calculations from Fig. 6(a)~(f), except that the number of quantum state measurements is set to 2 in all cases in Fig. 7. From the figures, it is clear that all peaks exceeding the $3\sigma$ threshold correspond to eigenstates. It is worth noting that some eigenstates inherently satisfy $|c_i|^2 > 3\sigma$, but due to spectral leakage, their amplitudes $a_i < 3\sigma < |c_i|^2$, leading to the failure to identify these eigenstates.

To identify eigenstates with smaller $|c_i|^2$, a more efficient approach than increasing the number of quantum measurements is to select a different reference state. For example, in the case of the $H_4$ molecular chain shown in Fig. 6(g), the reference state is the HF state, with $N = 5714$. Even with only one quantum measurement ($M = 1$, where the possible values of $\bar{q}(n)$ are limited to -1 and 1), the threshold is $3\sigma = 0.112$. Since the ground state component overwhelmingly dominates in the HF state, the overlap between the two wavefunctions is nearly 1, making the ground state easily detectable—though only the ground state is obtained. To compute some of the lower-lying excited states, one can switch the reference state to a single or double excitation configuration, as these states typically have substantial overlap with low-energy excited states without significantly increasing the quantum circuit cost. The results are shown in Fig. 8, where it is evident that, even with just one quantum measurement, most of the low-energy states can be identified as long as an appropriate reference state is selected.

### 3.6 Analysis of the Computational Cost of the XZ24 Method

In this section, we analyze the quantum resource requirements of the XZ24 method for calculating a system of size $L$ (representing either the number of atoms in a molecular system or lattice sites in a condensed matter model) with a target absolute accuracy of $\varepsilon$. The quantum resources considered include the number of qubits, quantum gates, and quantum measurements. Besides one auxiliary qubit, the computation necessitates a set of quantum registers to encode the initial reference state. In atomic and molecular systems, the required number of qubits depends on the configuration numbers derived from HF calculations, which in turn is influenced by the chosen basis set but generally scales with the number of atoms. For systems modeled by the Heisenberg or Ising models, the number of qubits corresponds directly to the number of lattice sites. Consequently, the storage complexity of the XZ24 method is $O(L)$.

As shown in Figure 1, the primary quantum circuit cost in the XZ24 method arises from implementing the single-qubit-controlled time evolution operator $e^{-i\hat{H}t}$. To compile this operator, a Trotter expansion is employed to decompose it into individual $e^{-i\hat{h}_i t}$ operators, which are compiled separately. Therefore, the circuit complexity is predominantly determined by the number of Pauli strings $\hat{h}_i$ in the Hamiltonian $\hat{H}$, denoted as $N_T$. The relationship between $N_T$ and the system size $L$ varies across different systems. For instance, in the one-dimensional Heisenberg model, the number of $\hat{h}_i$ operators increases linearly with system size, yielding $N_T \propto L$. In contrast, for two- and three-dimensional Heisenberg models, we have $N_T \propto L^2$ and $N_T \propto L^3$, respectively. In atomic and molecular systems, the number of $\hat{h}_i$ terms is even larger, typically scaling as $N_T \propto L^4$. In this analysis, we assume the worst-case scenario, where $N_T \propto L^4$. Given that compiling each $e^{-i\hat{h}_i t}$ operator requires $O(L)$ basic quantum gates, the total number of quantum gates required by the XZ24 method is $O(L^5)$. It is important to note that this estimate represents the worst-case scenario. In practice, several methods have been proposed to reduce the execution cost of the $e^{-i\hat{H}t}$ operator, such as adaptive-step Trotterization [65] and tensor network optimizations [64,67]. A valuable direction for future research is exploring the application of these techniques to optimize XZ24 circuits, addressing the practical challenges posed by decoherence and circuit noise on quantum computers.

Regarding the circuit sampling cost, as discussed in Section 3.2, the maximum sampling interval $\Delta_{\max} = \frac{\pi}{|E_0|} \propto 1/L$. From the analysis in Section 3.4, the maximum evolution time $T_{\max}$ determines the eigen-energy quantization unit $\delta = 2\pi/T_{\max}$, which, in turn, defines the calculation accuracy. To ensure that the absolute accuracy is less than $\varepsilon$, the maximum evolution time is $T_{\max} = 2\pi/\varepsilon$, leading to the total number of evolution time samples $N = T_{\max}/\Delta_{\max} = 2|E_0|/\varepsilon \propto L/\varepsilon$. For each evolution time point, determining the value of $q(n)$ requires multiple ($M$) circuit samples. However, as discussed in Section 3.5, $M$ does not affect the accuracy of the final result and does not scale with system size. Instead, $M$ only impacts the identification of eigenstate spectral peaks. If an appropriate reference state is chosen such that the eigenstates of interest have significant overlap with the reference state, $M$ can be kept small. In summary, the total circuit sampling cost is $N \cdot M = O(L \cdot \varepsilon^{-1})$. It is worth noting that if only relative accuracy ($\epsilon$) is required, instead of absolute accuracy, then, according to Eq. (19) in Section 3.4, $N = 2/\epsilon$, and the total number of circuit samples becomes $N \cdot M = O(\epsilon^{-1})$, which remains independent of system size.

## 4. Conclusions

In this paper, we propose an efficient XZ24 quantum algorithm specifically designed for calculating the eigenstates of quantum systems. We extensively explore both the theoretical foundations and practical applications of the XZ24 algorithm, supporting its effectiveness and reliability through numerical simulations on various quantum systems. Our results demonstrate that the XZ24 algorithm can accurately compute eigen-energies within chemical accuracy, offering a promising approach for quantum simulations. Furthermore, we analyze the computational errors and spectral characteristics of the XZ24 algorithm, emphasizing the significance of parameters such as the maximum evolution time and sampling interval. Numerical experiments confirm the algorithm's high precision in identifying eigenstates, and we discuss strategies for determining the sign of eigenvalues. When the sampling interval satisfies the Nyquist sampling theorem, the computational accuracy of the algorithm is strongly influenced by the maximum evolution time, while the impact of the sampling interval on result accuracy is minimal. Consequently, it is feasible to increase the sampling interval to reduce the number of required samples without compromising accuracy. When executing the algorithm, it is crucial to first define the desired accuracy, such as chemical accuracy, and use this to determine the maximum evolution time. Classical approximation methods can aid in identifying a reference state that closely approximates the target eigenstate and in estimating the range of its eigenenergy, thereby providing an estimate for the maximum frequency of the $q(n)$ signal. This, in turn, helps determine the optimal sampling interval and minimizes the number of measurements needed for each sampling of $q(n)$, thereby reducing quantum sampling costs while maintaining computational precision. Compared to existing quantum algorithms like the VQE, the main disadvantage of the XZ24 method lies in its greater circuit depth and higher demand for quantum error correction. However, its primary advantage is the lower sampling cost and the ability to compute multiple eigenvalues simultaneously, which enhances computational efficiency. Moreover, the VQE method demands significant resources for eigenstate preparation, with no guarantee of success, as the design and optimization of the ansatz circuit pose considerable challenges. In contrast, the XZ24 method requires only the preparation of a reference state on the quantum register that exhibits a non-negligible overlap with the target eigenstate, which is often much easier to achieve. With advancements in quantum computing hardware, challenges such as quantum decoherence and circuit noise are expected to diminish. In such a scenario, the XZ24 method, with its simplicity, cost-effectiveness, and capability to handle multiple eigenstates, will offer significant value in the quantum computing landscape. Future research should explore its applicability to more complex systems and evaluate its performance in real-world applications.

**Acknowledgments**

This work is supported in part by the National Natural Science Foundation of China (Grant No. 22273069).